# Atomically Thin CrCl$_3$: An in-Plane Layered Antiferromagnetic Insulator

Xinghan Cai,[1,2] Tiancheng Song,[1] Nathan P. Wilson,[1] Genevieve Clark,[3] Minhao He,[1] Xiaoou Zhang,[4] Takashi Taniguchi,[5] Kenji Watanabe,[5] Wang Yao,[6] Di Xiao,[4] Michael A. McGuire,[7] David H. Cobden,[1] Xiaodong Xu[1,3]*

[1]Department of Physics, University of Washington, Seattle, Washington 98195, USA.
[2]National Key Laboratory of Science and Technology on Micro/Nano Fabrication, Department of Micro/Nano Electronics, Shanghai Jiao Tong University, Shanghai 200240, China.
[3]Department of Materials Science and Engineering, University of Washington, Seattle, Washington 98195, USA.
[4]Department of Physics, Carnegie Mellon University, Pittsburgh, Pennsylvania 15213, USA.
[5]National Institute for Materials Science, Tsukuba, Ibaraki 305-0044, Japan.
[6]Department of Physics and Center of Theoretical and Computational Physics, University of Hong Kong, Hong Kong, China.
[7]Materials Science and Technology Division, Oak Ridge National Laboratory, Oak Ridge, Tennessee 37831, USA.

*Correspondence to: xuxd@uw.edu

**Abstract:** The recent discovery of magnetism in atomically thin layers of van der Waals (vdW) crystals has created new opportunities for exploring magnetic phenomena in the two-dimensional (2D) limit. In most 2D magnets studied to date the c-axis is an easy axis, so that at zero applied field the polarization of each layer is perpendicular to the plane. Here, we demonstrate that atomically thin CrCl$_3$ is a layered antiferromagnetic insulator with an easy-plane normal to the c-axis, that is the polarization is in the plane of each layer and has no preferred direction within it. Ligand field photoluminescence at 870 nm is observed down to the monolayer limit, demonstrating its insulating properties. We investigate the in-plane magnetic order using tunneling magnetoresistance in graphene/CrCl$_3$/graphene tunnel junctions, establishing that the interlayer coupling is antiferromagnetic down to the bilayer. From the temperature dependence of the magnetoresistance we obtain an effective magnetic phase diagram for the bilayer. Our result shows that CrCl$_3$ should be useful for studying the physics of 2D phase transitions and for making new kinds of vdW spintronic devices.

The experimental observation of magnetism in atomically thin films[1-8] has opened a new avenue to study novel magnetic multilayered devices using 2D materials[9-21]. To date, all magnetic compounds that have been successfully synthesized down to the monolayer limit exhibit strong Ising anisotropy, which supports an out-of-plane magnetization within each layer. In this regard,

CrCl$_3$ represents a curious case in that bulk CrCl$_3$ is known to be a weakly coupled, layered antiferromagnet with small XY anisotropy (Figure 1a)[22-25]. Therefore, exfoliating CrCl$_3$ offers the intriguing possibility to realize a strictly 2D magnetic insulator with in-plane magnetization. This could be useful for investigating how a magnet with weak anisotropy evolves on approaching the 2D limit. The monolayer may behave as a good approximation to the 2D XY-model, analogous to thin film superfluids, superconductors and liquid crystals[26,27], allowing the study of the Berezinskii–Kosterlitz–Thouless (BKT) transition[28] and related phenomena. Moreover, the in-plane magnetization is also complementary to the out-of-plane magnetization in terms of the proximity effect[29-33] when CrCl$_3$ is layered with other 2D materials.

CrCl$_3$ flakes were exfoliated onto SiO$_2$/Si substrates. In atomic force microscope images (Figure 1b), bilayers and trilayer regions appear 1.2 nm and 1.8 nm thick, respectively, consistent with a previous report[25]. Figure 1c shows both the circular polarization components of the photoluminescence (PL) from a ~ 5 nm thick sample at a temperature of 2 K, well below the Neel temperature of ~ 14 K. The excitation laser is 20 µW, linearly polarized, and at 532 nm. A single broad peak centered at ~ 870 nm is seen, analogous to the parity-forbidden *d-d* ligand-field transition in CrI$_3$[34]. Unlike in CrI$_3$, however, the two polarization components are indistinguishable, indicating that there is no out-of-plane magnetization component in CrCl$_3$. This PL emission feature persists down to the monolayer CrCl$_3$ (Figure 1d), implying that the insulating nature retains in CrCl$_3$ to its 2D limit.

To investigate the in-plane magnetization we employ vertical tunneling measurements, which are sensitive to the relative alignment of spins in different layers. As sketched in Figure 2a, a few layer CrCl$_3$ tunnel barrier is sandwiched between top and bottom thin graphite flakes to form magnetic tunnel junction (MTJ), this being further encapsulated in hexagonal boron nitride (h-BN) to avoid degradation (see Methods). Figure 2b shows the tunneling current, $I_t$, as a function of DC bias voltage, $V$, measured at 2 K for a bilayer CrCl$_3$ device whose optical image is inset. With an applied in-plane magnetic field $\mu_0 H_{in}$=2 T (red trace) the current is enhanced relative to its value at zero field (blue trace), implying a change in the spin configuration of the CrCl$_3$[35]. The bottom inset shows the tunneling magnetoresistance (TMR), defined as $100\% \times [I_t(2\text{ T}) - I_t(0)]/I_t(0)$[13]. The TMR is largest at low bias, where the response is linear. Figure 2c shows the tunneling current, $I_t$, here measured using a small AC bias of $V_{ac} = 1$ mV, as a function of magnetic field. For either in-plane field ($H_{in}$, orange) or out-of-plane field ($H_{out}$, green), $I_t$ increases smoothly before plateauing out when the field exceeds a critical value ($\mu_0 H_{in}^c = 1.2$ T and $\mu_0 H_{out}^c = 1.5$ T). This behavior is similar to that seen for an in-plane field in CrI$_3$ MTJs. It implies that the spin polarizations of the two layers are opposite at zero field, suppressing the tunneling current by a spin-filtering effect, but become increasingly aligned due to the Zeeman term in an external field. This in turn suggests that the layered antiferromagnetic order of the bulk crystal, with in-plane ferromagnetic polarization that alternates in direction between adjacent layers, persists in the bilayer limit.

The fact that $H_{out}^c$ is only slightly larger than $H_{in}^c$, and the smooth rise of $I_t$ seen for both field orientations, are consistent with relatively weak anisotropy in the material. Indeed, the field dependence of the tunneling current can be well fit by a simple model of two antiferromagnetically coupled macrospins with easy-plane anisotropy,

$$H = J_{AF} \mathbf{s}_1 \cdot \mathbf{s}_2 + \frac{K}{2}\left(s_{1,z}^2 + s_{2,z}^2\right) - g\mu_B \mathbf{B} \cdot (\mathbf{s}_1 + \mathbf{s}_2),$$

where $J_{AF} > 0$ is the interlayer antiferromagnetic exchange, $K > 0$ is the easy-plane anisotropy energy, and $\mathbf{B}$ is the magnetic field. According to our fitting, the anisotropy is $KS/g\mu_B = 0.27$ T and interlayer coupling is $J_{AF}S/g\mu_B = 0.6$ T (See Supplementary material). We note that there is some asymmetry of $I_t$ as a function of $H_{in}$, associated with hysteresis as shown in Figure 2d. This might be due to a spin-flop transition associated with a previously reported, very small easy-axis anisotropy (10 Oe) within the easy-plane[24,25]. No hysteresis is seen as a function of $H_{out}$ (Figure 2d, inset), as expected for spin canting in a magnetic field perpendicular to the zero-field polarization. This is in sharp contrast to the situation in bilayer CrI3, where the transition from antiferromagnetic to a fully aligned ferromagnetic state in a perpendicular field is a sudden spin-flip transition.

The picture of layered antiferromagnetism is further supported by measurements on a trilayer CrCl3 MTJ device (Figure 3), an optical micrograph of which is inset to Figure 3b. The thicker barrier presented by a trilayer leads to much weaker tunneling which necessitates a much larger DC bias of 600 mV to achieve a suitable signal to noise ratio. As seen in Figure 3a, $I_t$ increases smoothly with $H_{out}$, again consistent with gradual spin canting out of the layers as in the bilayer case[13]. However, as a function of $H_{in}$ the current exhibits a plateau near zero field below a critical value, above which it smoothly rises and reaches a plateau when the magnetic field is large enough to align all three layers to the fully spin polarized state.

To test for magnetic anisotropy within the plane of the layers, we measured the tunneling current while varying the orientation, $\theta$, of $H_{in}$ within the plane. Figure 3b shows $I_t$ as a function of $H_{in}$ at selected angles, while Figure 3c is a polar plot of $I_t$ as a function of angle at selected magnetic fields. No significant dependence on $\theta$ is observed, implying that the overall magnetic order can rotate freely about the c-axis, i.e., the plane of the layers is an easy plane. The observed plateau of tunneling current as a function of in-plane magnetic field is due to the uncompensated total magnetization in a trilayer, consistent with the picture of layered antiferromagnetism. We conclude that atomically thin CrCl3 can be well described as a layered antiferromagnetic insulator with easy-plane magnetic anisotropy.

Figure 4a shows the tunneling current $I_t$ measured AC with $V_{ac} = 5$ mV as a function of $T$ at selected in-plane magnetic fields. At $H_{in} = 0$, on increasing $T$ from base, at first $I_t$ increases. This can be explained by suppression of the spin filtering effect due to thermal fluctuations which reduce the degree of antiferromagnetic alignment between the layers. Then, above a peak at the intralayer ferromagnetic critical temperature $T \sim 17$ K, $I_t$ starts to decrease. This suggests that ferromagnetic order within each layer assists coherent tunneling, and rising temperature suppresses this order. At still higher temperatures $I_t$ increases again, as expected since once no magnetic order remains the tunneling rate should simply be activated. When an in-plane field is applied it competes with the antiferromagnetic alignment and reduces the temperature at which the AFM order appears, so the peak moves to lower $T$. It also increases net spin alignment above the critical temperature and so increases the current. For $\mu_0H_{in} > 1.2$ T, the antiferromagnetism is completely suppressed and the spins are fully polarized at base temperature.

The results in Figure 4a can be presented in a more visualized way: $dI_t/dT$ has been plotted as a function of the temperature at selected $\mu_0H_{in}$ (Figure 4b). In the top panel, $dI_t/dT$ clearly exhibits two times of the sign flip as $T$ increases. This is similar to what has been reported in CrCl$_3$ 3D bulk crystal[25], which is interpreted as a two-step phase transition, i.e. from the layered antiferromagnetic phase to the ferromagnetic phase (first transition) and eventually to the paramagnetic phase (second transition) as $T$ increases. Here we assign the peak (directed by the green arrow) in each curve to the critical temperature of the first transition and the dip (directed by the blue arrow) to the second one. For magnetic fields higher than 1.2 T, as shown in the bottom panel, no local maximum shows up, thus only $T_c$ for the second transition is determined. In Figure 4c, we present the 2D semi-log plot of the normalized $dI_t/dT$ as a function of both $\mu_0H_{in}$ and $T$ (see supplementary material for the plot without the normalization of $dI_t/dT$). This plot, though not a legitimate phase diagram, facilities comparison with the phase diagram of the 3D bulk crystal in which distinct transitions are identified between the layered AFM phase, a fully spin-polarized phase, and paramagnetic phase[25].

To recap, we find that the van der Waals magnetic insulator CrCl$_3$ shows layered antiferromagnetism down to the bilayer limit with the magnetic moments in the plane of the layers and little or no anisotropy within the plane. CrCl$_3$ thus provides a new model system for studying 2D in-plane magnetism and its proximate coupling to other 2D materials, such as 2D superconductors, topological insulators, and stacked heterostructures[30,36] for potential use in novel magneto-electronic devices[13-18].

During the final preparation of the manuscript, we became aware of the two relevant works posted on ArXiv[37,38].

**METHODS**

**Device fabrication:** V/Au (4 nm/40 nm) metal electrodes were deposited on a 90 nm SiO$_2$/Si substrate in the vicinity of the exfoliated h-BN thin flake (bottom h-BN), following the standard e-beam lithography procedures. Bulk CrCl$_3$ crystals were grown by chemical vapour transport, as described in detail in ref[25]. Few layer CrCl$_3$, cleaved from these crystals in an inert gas glovebox with oxygen and water levels below 1 ppm, was stacked between the pre-exfoliated top h-BN/graphite and bottom graphite and transferred onto the bottom h-BN using a polymer-based dry transfer technique[39]. Top/bottom graphite flakes were contacted by the pre-patterned metal electrodes and the tunnel junction was then fully encapsulated before finally dissolving the polymer in chloroform outside the glovebox. The thickness of the h-BN (5-35 nm) and graphite (2-8 nm) were determined by atomic force microscopy, while the number of layers of CrCl$_3$ was identified from the optical contrast.

**Optical measurements:** Low-temperature magneto-PL measurements were performed in a closed-cycle cryostat with a superconducting magnet with field normal to the sample. The sample was excited by a continuous-wave laser (633 nm or 532 nm) with the power below 50 μW to avoid sample heating and degradation. The PL signal was detected by a silicon CCD array and the excitation and detection polarization were controlled using linear polarizers and half- and quarter-wave plates.

**Electrical measurements**: All magneto-transport measurements were carried out in a PPMS DynaCool cryostat (Quantum Design, Inc) with a base temperature of 1.8 K and magnetic field up to 9 T. The $CrCl_3$ MTJ devices were mounted in a Horizontal Rotator probe using either a Rotator Universal Sample Board (P103C), which allows rotations about an axis in the sample plane perpendicular to the magnetic field, or a Rotator Field Parallel Sample Board (P103D) for rotation about the magnetic field axis. The bias voltage is applied to the top graphite contact while the current from the bottom graphite contact is measured using a virtual-earth current preamplifier (DL Instruments; Model 1211). AC measurements were made using a lock-in amplifier (Stanford Research 830) at 13.7 Hz.

**Figure Captions:**

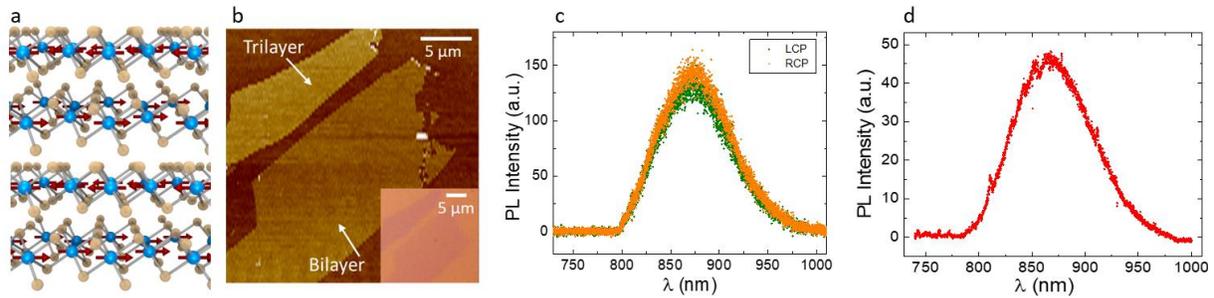

**Figure 1. $CrCl_3$ crystal structure and the optical characterization.** (a) Side view of the structure of $CrCl_3$. The colored balls represent Cr (blue) and Cl (grey), respectively. Red arrows indicate the magnetization. (b) Atomic force micrograph of few layer $CrCl_3$ flakes. The inset is an optical micrograph of the same area. (c) Left and right circular polarization components of photoluminescence from a 5 nm thick flake under a linearly polarized excitation at 532 nm. (d) Photoluminescence spectrum of a monolayer $CrCl_3$.

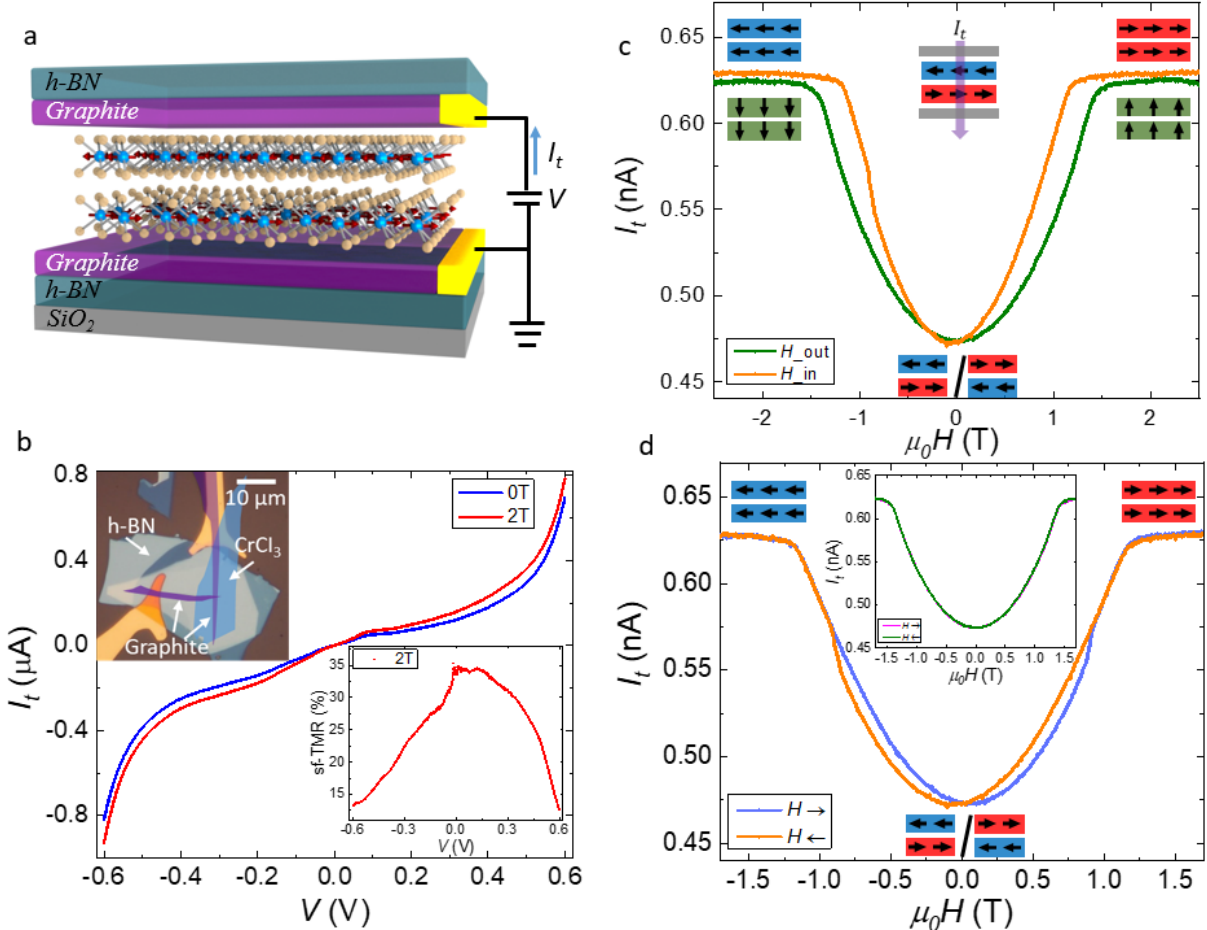

**Figure 2. Magnetic field dependent tunneling in a bilayer CrCl₃ spin-filter magnetic tunnel junction (sf-MTJ).** (a) Schematic of device with graphite contacts encapsulated by h-BN. (b) Tunneling current as a function of the DC bias voltage with (red) and without (blue) an in-plane magnetic field $\mu_0H_{in}$=2 T. Top inset: optical microscope image of the device with false coloring for clarity. Bottom inset: TMR ratio. (c) Tunneling current $I_t$ as a function of in-plane (orange) and out-of-plane (green) magnetic field for bias $V_{ac}$ = 1 mV. The corresponding magnetic states are indicated. (d) Tunneling current vs in-plane magnetic field sweep in both directions. The magnetic configurations are indicated. Top inset: Tunneling current vs out-of-plane field swept in both directions, showing no hysteresis.

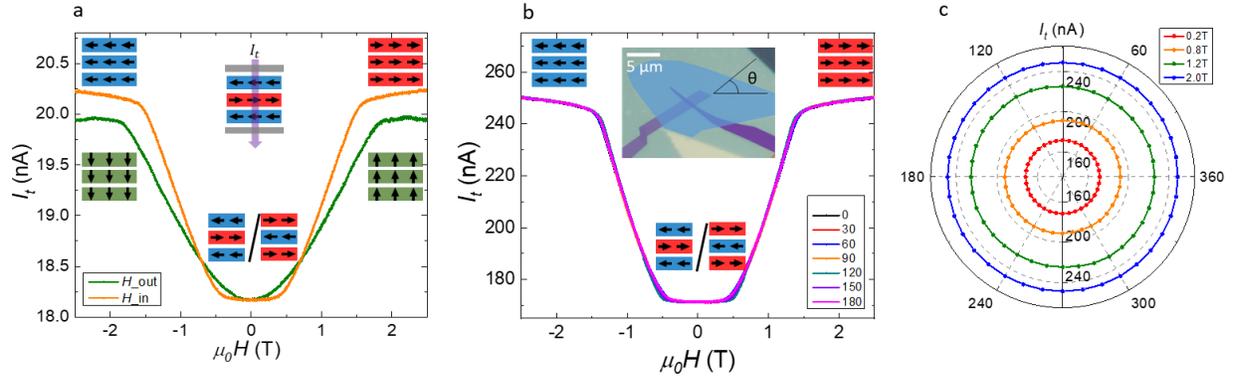

**Figure 3. Magnetic field dependent tunneling in a tri-layer CrCl₃ sf-MTJ device.** (a) Tunneling current $I_t$ as a function of the in-plane (orange) and out-of-plane (green) magnetic field at $V_{dc} = 600$ mV. The corresponding magnetic states are indicated. (b) Tunneling current as a function of in-plane magnetic field at selected orientations $\theta$ (here $V_{dc} = 750$ mV). The inset is a false-color optical micrograph of the device with $\theta$ defined. (c) Polar plot of the tunneling current as a function of $\theta$ at selected magnetic fields.

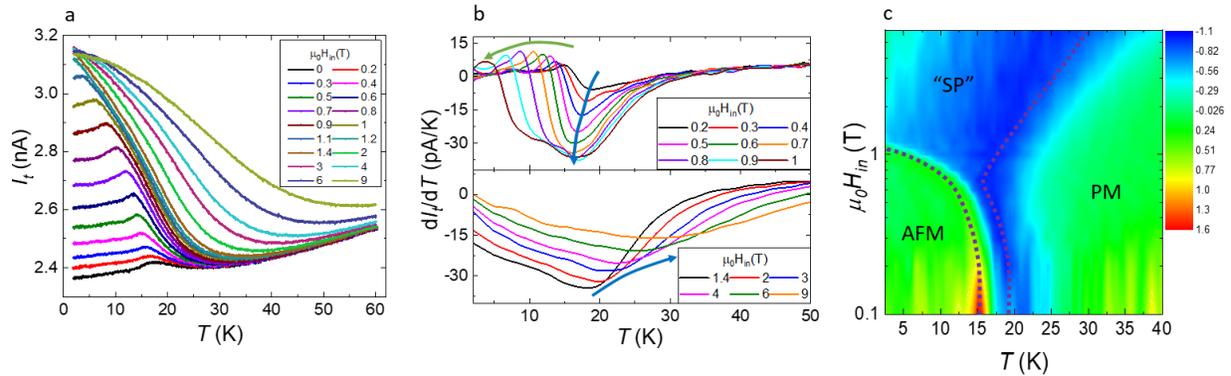

**Figure 4. Temperature dependent tunneling and the magnetic phase diagram of bilayer CrCl₃.** (a) Tunneling current $I_t$ as a function of the temperature at selected in-plane magnetic fields with the AC bias voltage $V_{ac} = 5$ mV. (b) The derivative of the tunneling current to the temperature at low (top panel) and high (bottom panel) in-plane magnetic fields. The green/blue arrows indicate the critical temperature of the transition from layered antiferromagnetic/spin polarized to spin polarized/paramagnetic states, respectively. (c) Color plot of normalized $dI_t/dT$ as a function of in-plane magnetic field $H_{in}$ and temperature $T$. AFM: Layered antiferromagnetic. PM: Paramagnetic. SP: Fully spin polarized.

**Acknowledgements:** Funding: The work at UW is mainly supported by NSF MRSEC 1719797. Device fabrication and transport measurements are partially supported by NSF-DMR-1708419. Photoluminescence measurement is supported by DOE BES DE-SC0012509. Work at HKU is supported by the Croucher Foundation (Croucher Innovation Award), RGC of HKSAR (17303518P). Work at ORNL (M.A.M.) was supported by the US Department of Energy, Office of Science, Basic Energy Sciences, Materials Sciences and Engineering Division. K.W. and T.T. acknowledge support from the Elemental Strategy Initiative conducted by the MEXT, Japan, A3 Foresight by JSPS and the CREST(JPMJCR15F3), JST. D.X. acknowledges the support of a Cottrell Scholar Award. X.X. acknowledges the support from the State of Washington funded Clean Energy Institute and from the Boeing Distinguished Professorship in Physics.

**Author Contributions:** X.X., X.C., M.A.M. conceived the project. X.C. fabricated the devices, performed the experiments, and analyzed the data, assisted by T.S. and M.H. N.P.W. and G.C. performed photoluminescence measurements. X.Z. analyzed the macrospin model and the tunneling current dependence on the magnetic structure. X.X., D.X., W.Y., and D.H.C. supervised the project. M.A.M. provided and characterized bulk $CrCl_3$ crystals. T.T. and K.W. provided and characterized bulk hBN crystals. X.C., X.X., D.H.C. and D.X. wrote the manuscript with input from all authors.

**Competing Financial Interests:** The authors declare no competing financial interests.

**Data Availability**: The data that support the findings of this study are available from the corresponding author upon reasonable request.

Supplementary Materials for

# Atomically-Thin CrCl₃: An in-Plane Layered Antiferromagnetic Insulator

Xinghan Cai,[1,2] Tiancheng Song,[1] Nathan P. Wilson,[1] Genevieve Clark,[3] Minhao He,[1] Xiaoou Zhang,[4] Takashi Taniguchi,[5] Kenji Watanabe,[5] Wang Yao,[6] Di Xiao,[4] Michael A. McGuire,[7] David H. Cobden,[1] Xiaodong Xu[1,3]*

[1]Department of Physics, University of Washington, Seattle, Washington 98195, USA.
[2]National Key Laboratory of Science and Technology on Micro/Nano Fabrication, Department of Micro/Nano Electronics, Shanghai Jiao Tong University, Shanghai 200240, China.
[3]Department of Materials Science and Engineering, University of Washington, Seattle, Washington 98195, USA.
[4]Department of Physics, Carnegie Mellon University, Pittsburgh, Pennsylvania 15213, USA.
[5]National Institute for Materials Science, Tsukuba, Ibaraki 305-0044, Japan.
[6]Department of Physics and Center of Theoretical and Computational Physics, University of Hong Kong, Hong Kong, China.
[7]Materials Science and Technology Division, Oak Ridge National Laboratory, Oak Ridge, Tennessee 37831, USA.

**S1. $I_t - V$ characterization with out-of-plane magnetic field in the bilayer CrCl₃ MTJ.**

The tunneling current $I_t$ at selected $\mu_0 H_{out}$ and the extracted sf-TMR ratio as a function of the DC bias voltage are shown in Figures S1a&b, respectively. $I_t$ increases linearly at low bias voltage and exponentially at high bias voltage, which is consistent with Figure 2b. As shown in Figure S1b, the TMR at high $\mu_0 H_{out}$ exhibits oscillating behavior in the low bias regime and becomes negative at large $V$. This is possibly related to the Landau level formation in the few layer graphite electrodes induced by $\mu_0 H_{out}$, which modulates the carrier density in top and bottom contacts, resulting in a change of the tunneling resistance. Quantitative analysis requires more detailed information about the band structure of CrCl₃ tunnel barrier, and the thickness and doping dependent carrier density in graphite electrodes, which will not be discussed in this work.

**S2. Temperature dependent tunneling with out-of-plane magnetic field in the bilayer CrCl₃ MTJ.**

Figure S2 exhibits $I_t$ versus $T$ at selected out-of-plane magnetic fields for $V_{ac}$ = 5 mV. For small $\mu_0 H_{out}$, $I_t$ first increases with the temperature due to the gradually relaxed interlayer antiferromagnetic coupling of the magnetic moments, followed by a decrease owing to the vanishing intralayer ferromagnetic order, which is consistent with the two-stage magnetic phase transitions with increasing temperature in CrCl₃ as reported in the main text. The critical temperature corresponding to the first transition (antiferromagnetic state to spin polarized state) shifts to the lower temperature side with increasing $\mu_0 H_{out}$ as a result of the competition between the Zeeman energy and the interlayer coupling energy. The critical field ($\mu_0 H_{out}^c$ = 1.5 T), at which the magnetic moments get fully polarized at the base temperature, is larger than $\mu_0 H_{in}^c$ (1.2

T) shown in Figures 4a-c, which further confirms the in-plane magnetization of bilayer $CrCl_3$ as demonstrated in the manuscript.

## S3. Bias dependent tunneling in the bilayer $CrCl_3$ MTJ.

The tunneling current as a function of $\mu_0 H_{in}$ and $\mu_0 H_{out}$ at selected DC bias voltages are displayed in Figures S3a and b, respectively. Both $\mu_0 H_{in}^c$ and $\mu_0 H_{out}^c$ are independent of $V$ in the measured bias voltage range. In the tunneling regime ($V = 5$ mV), the oscillating behavior of $I_t$ at large $\mu_0 H_{out}$ shown with the black curve in Figure S3b is attributed to the quantum oscillations in top and bottom graphite contacts. For $V = 550$ mV, $I_t$ decreases significantly when a large out-of-plane magnetic field is applied, which is possibly related to the Landau level formation and density of states redistribution in graphite electrodes considering that $I_t$ remains almost unchanged at large $\mu_0 H_{in}$.

## S4. Effective magnetic phase diagram of the bilayer $CrCl_3$.

The 2D semi-log plot of $dI_t/dT$ as a function of $\mu_0 H_{in}$ and $T$ of the bilayer $CrCl_3$ MTJ is shown in Figure S4. $dI_t/dT$ is extracted from Figure 4a. The antiferromagnetic to spin-polarized state and the spin-polarized to paramagnetic state transitions are assigned to occur at the local maximum and minimum in the $dI_t/dT - T$ curves as shown in Figure 4b. In the main text, Figure S4 is replotted as Figure 4c by normalizing each $dI_t/dT - T$ curve by the value of $dI_t/dT$ at its local minimum, in order to exhibit the three distinct magnetic states in a more visualized way.

## S5. Electron tunneling through multilayer $CrCl_3$

We have explored the electron tunneling through multilayer $CrCl_3$ (more than 3 but less than 10 layers) and show the results of two typical devices in Figures S5a-c and Figures S5d-f respectively. As illustrated in Figures S5a and d, the tunneling is much weaker in both devices compared to the bilayer sample in the main text, but the TMR ratio gets enhanced, owing to the increased $CrCl_3$ barrier thickness. For the in-plane/out-of-plane magnetic field sweep, while the regular spin canting behavior is observed in the first device (see Figure S5b and S5c), the second device shows an additional dip/kink when sweeping the field across $\mu_0 H_{1in}^c \sim 0.25$ T/$\mu_0 H_{1out}^c \sim 0.6$ T as shown in Figure. 5e and its inset. Whether these extra features are due to the inhomogeneity of the tunnel barrier is still to be understood, but the fact that $\mu_0 H_{1in}^c$ and $\mu_0 H_{1out}^c$ are close to the critical field at which the ferromagnetic-like spin configuration in bulk $CrCl_3$ is realized may suggest that some of the multilayer $CrCl_3$ MTJs recover the properties of the bulk crystal, associated with a dimensional crossover, which is also observed in few layer $Fe_3GeTe_2$[1] and other ultrathin magnetic systems[2]. Figure S5f shows that this low field feature of $I_t$ exists at all angles between $\mu_0 H_{in}$ and the device. The red curve in the inset's polar plot indicates that at very low field, the tunneling current is slightly anisotropic. This is possibly related to a spin-flop transition, which has been reported in bulk $CrCl_3$[3, 4] with an in-plane magnetic field around $\mu_0H_{flop}$ ~ 160 Oe and will be discussed more in detail in the next section.

## S6. Spin-flop transition in multilayer $CrCl_3$.

The magnetic anisotropy of the bulk $CrCl_3$ is weak with the estimated in-plane anisotropy field of only ~10 Oe. A spin-flop transition is expected to occur when an in-plane magnetic field is applied parallel to the spins. Experimentally, some signature of the spin-flop has been observed at $\mu_0H_{in}$ = 100 ~ 200 Oe in $CrCl_3$ bulk crystals[3, 4]. However, since the samples used in previous magnetization or magneto-optical measurements usually consist of several $CrCl_3$ thin flakes with magnetic domains, which are not co-aligned with respect to the same spin orientation, both the spin-flop and the coherent rotation (spin-canting) occur when an in-plane magnetic field is applied, thus any sharp features of the transition are smoothed due to the averaging effect.

In our work, the stacked heterostructure is composed of single crystal $CrCl_3$ flake and the tunnel junction area is usually less than ~1 µm$^2$ to avoid the averaging effect caused by lateral magnetic domain structures. According to the results shown in the main text, no obvious signature of a spin-flop around $\mu_0H_{in}$ ~ 160 Oe has been observed in our bilayer, trilayer and the first multilayer (Figure S5a-c) $CrCl_3$ MTJs, implying that the magnetic properties of atomically thin $CrCl_3$ differ from the bulk crystal. On the other hand, Figure S6 demonstrates a possible spin-flop transition in the second multilayer device (Figure S5d-f) by sweeping $\mu_0H_{in}$ at different angles ($\theta$) in the low field regime. At $\theta$ = 80°, $I_t$ stays almost unchanged before $\mu_0H_{in}$ reaches a critical value, where a sharp jump of $I_t$ occurs, followed by a continuous up-turn upon further increasing $\mu_0H_{in}$. This is consistent with the spin-flop transition in a weakly anisotropic antiferromagnetic system, where the magnetization of the material suddenly rotates to the direction perpendicular to the magnetic field that is applied to its easy direction, and then the magnetic moments are gradually aligned by the continuously increasing field. A critical field of ~0.02T here is close to the spin-flop field of the bulk $CrCl_3$ crystal. When rotating the angle between $\mu_0H_{in}$ and the device, the sharp jump of $I_t$ gets smoothed, corresponding to a mixture of the spin-flop and the coherent rotation. As shown in the figure, at $\theta$ = 160°, the spin-flop almost disappears and is replaced by a continuous increase of $I_t$, which is interpreted in terms of the spin-canting effect.

## S7. Power dependent photoluminescence (PL) of monolayer $CrCl_3$

We investigate the insulating nature of the monolayer $CrCl_3$ by the power dependent PL measurement. Figure S7a shows the optical and atomic force micrograph of typical monolayer $CrCl_3$ samples. The single broad PL peak around 870 nm characterized using a linearly polarized laser excitation at 2.33 eV with selected powers is illustrated in Figure S7b, which we attribute to a ligand-field transition[5] and is consistent with the observation shown in Figure 1d in the main text. The PL intensity at 870 nm extracted from Figure S7b is plotted as a function of the laser power in Figure S7c, which can be well fitted with a straight line, indicating that there is no nonlinear effect in the excitation power regime up to 50 µW.

## S8. Modelling the tunneling current in the bilayer CrCl₃ MTJ

We use a simple model to fit the magnetic field dependent tunneling current in the bilayer CrCl$_3$. With a fixed bias voltage, the tunneling current is determined by two factors: the resistance within each layer and the tunneling probability between the two layers. Both of them depend on the directions of the magnetization of the two layers, which can be obtained by minimizing the spin Hamiltonian

$$H = J_{AF} \mathbf{s}_1 \cdot \mathbf{s}_2 + \frac{K}{2}\left(s_{1,z}^2 + s_{2,z}^2\right) - g\mu_B \mathbf{B} \cdot (\mathbf{s}_1 + \mathbf{s}_2),$$

where $J_{AF}$ is the interlayer antiferromagnetic exchange energy, $K$ is the easy plane anisotropy energy, and $\mathbf{B}$ is the magnetic field. Here we have assumed that the in-plane ferromagnetic exchange energy is large enough to align all the spins in each layer toward the same direction. The spin in each layer is represented by $\mathbf{s}_1$ and $\mathbf{s}_2$.

Given the magnetization direction, we use a similar method as mentioned in the supplementary material of Ref.[6] to calculate the tunneling current. Using symmetry analysis, the angular dependent resistance in each layer has the following form

$$\rho(\theta_{Mi}) = \rho_0 + \rho_1 \cos(2\theta_{Mi}),$$

where $\theta_{Mi}$ is the angle between the magnetization direction of the *i*-th layer and the out-of-plane z direction, and the parameters $\rho_0, \rho_1$ can be determined by fitting the experiment data. Now consider the tunneling probability between the two layers. If the magnetic field is along the z-direction, the tunneling probability is given by Ref. (6)

$$t(\theta_{M1}, \theta_{M2}) = \cos^2\frac{\theta_{M1} - \theta_{M2}}{2} t_{\parallel} + \sin^2\frac{\theta_{M1} - \theta_{M2}}{2} t_{\perp},$$

where $t_{\parallel}/t_{\perp}$ is the tunneling probability when the spins in the two layers are parallel/anti-parallel to each other, respectively. Therefore, the tunneling current is

$$\sigma(\theta_{M1}, \theta_{M2}) = const \times \frac{t(\theta_{M1}, \theta_{M2})}{\rho(\theta_{M1}) + \rho(\theta_{M2})}.$$

If the magnetic field is in-plane, then the magnetization is also in-plane. In this case, we simply regard the direction of the magnetic field as the new z-axis and replace $\theta_{Mi}$ with the in-plane polar angle.

We use this model to fit the in-plane and the out-of-plane tunneling current as shown in Figure S8. According to the fitting, the exchange interaction is $J_{AF}S/g\mu_B = 0.6$ T and $KS/g\mu_B = 0.27$ T.

**Figures:**

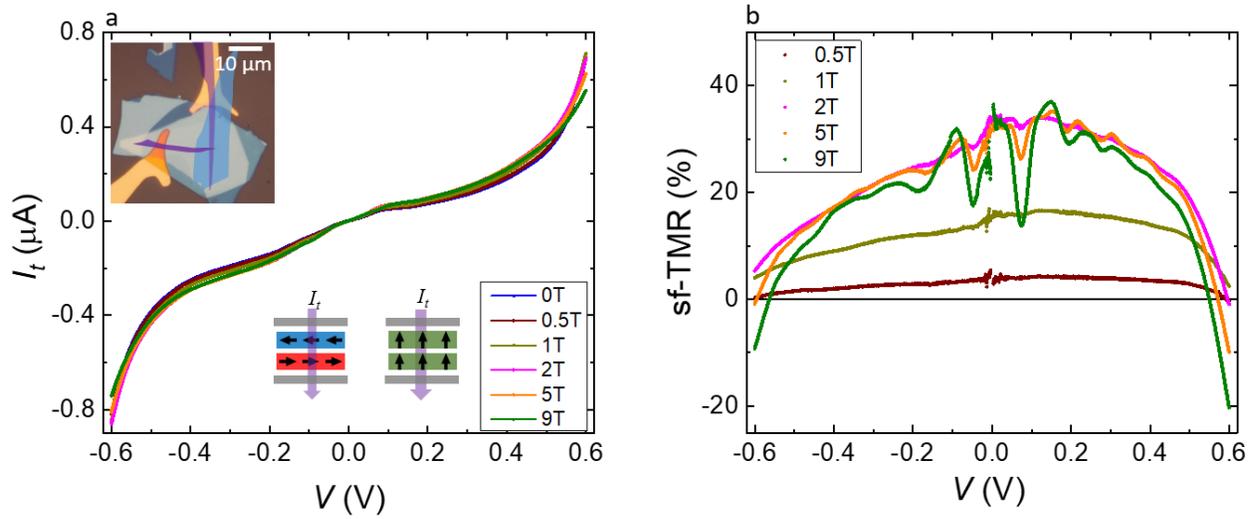

**Figure S1.** DC transport characterization of a bilayer CrCl$_3$ sf-MTJ with the out-of-plane magnetic field. (a) Tunneling current as a function of the DC bias voltage at selected out-of-plane magnetic fields. Top left inset: false color optical microscope image of the device. The bottom inset shows the corresponding magnetic states. (b) Extracted sf-TMR ratio as a function of the bias voltage based on (a).

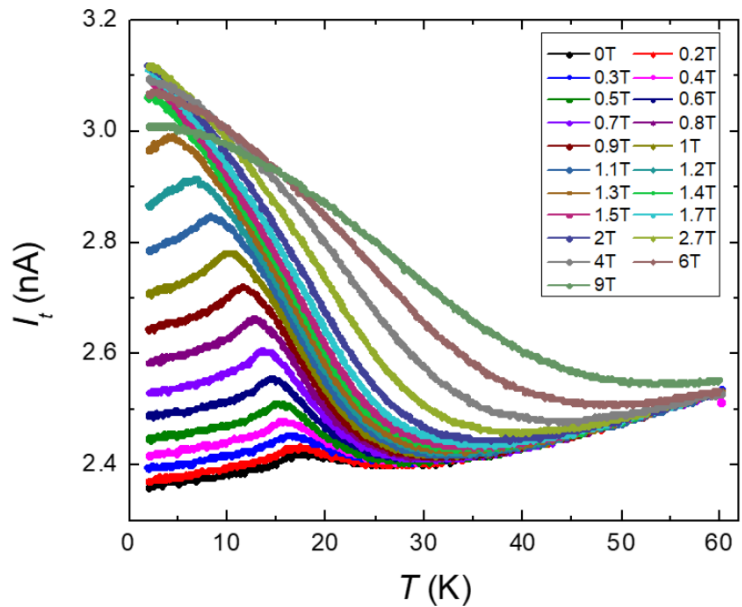

**Figure S2.** Temperature dependent tunneling of the bilayer CrCl$_3$ sf-MTJ with the out-of-plane magnetic field. Tunneling current as a function of the temperature at selected out-of-plane magnetic fields with the AC bias voltage $V_{ac}$ = 5 mV.

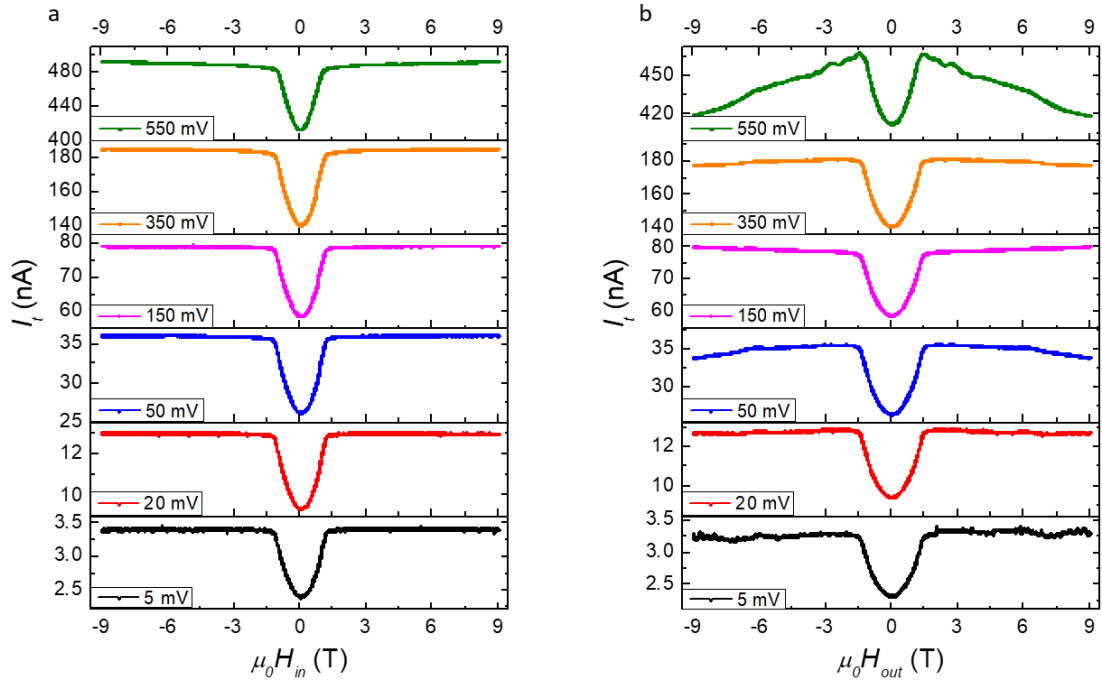

**Figure S3.** Magnetic field dependent tunneling of the bilayer CrCl$_3$ sf-MTJ. (a, b) Tunneling current as a function of the in-plane (a) and out-of-plane (b) magnetic field at selected DC bias voltage, respectively.

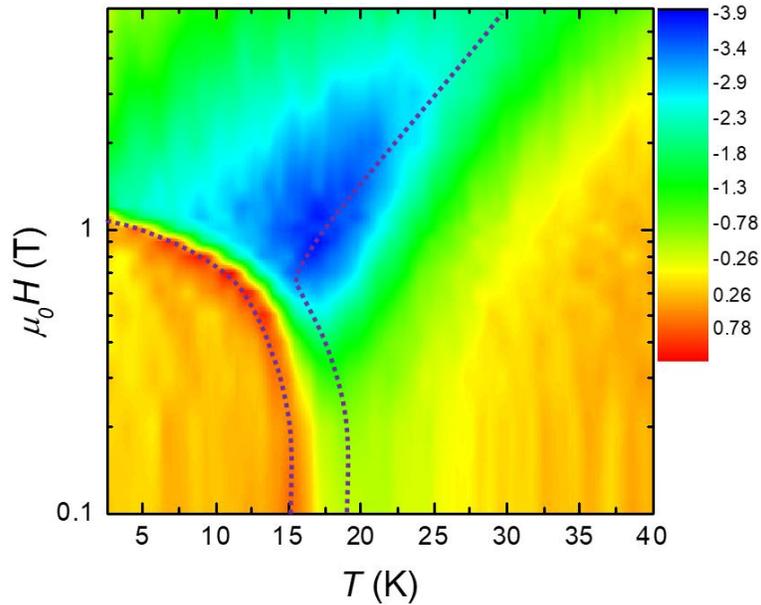

**Figure S4.** Effective magnetic phase diagram of the bilayer CrCl$_3$. 2D semi-log plot of d$I_t$/d$T$ as a function of the in-plane magnetic field and the temperature. Different magnetic phases are separated by purple dotted lines.

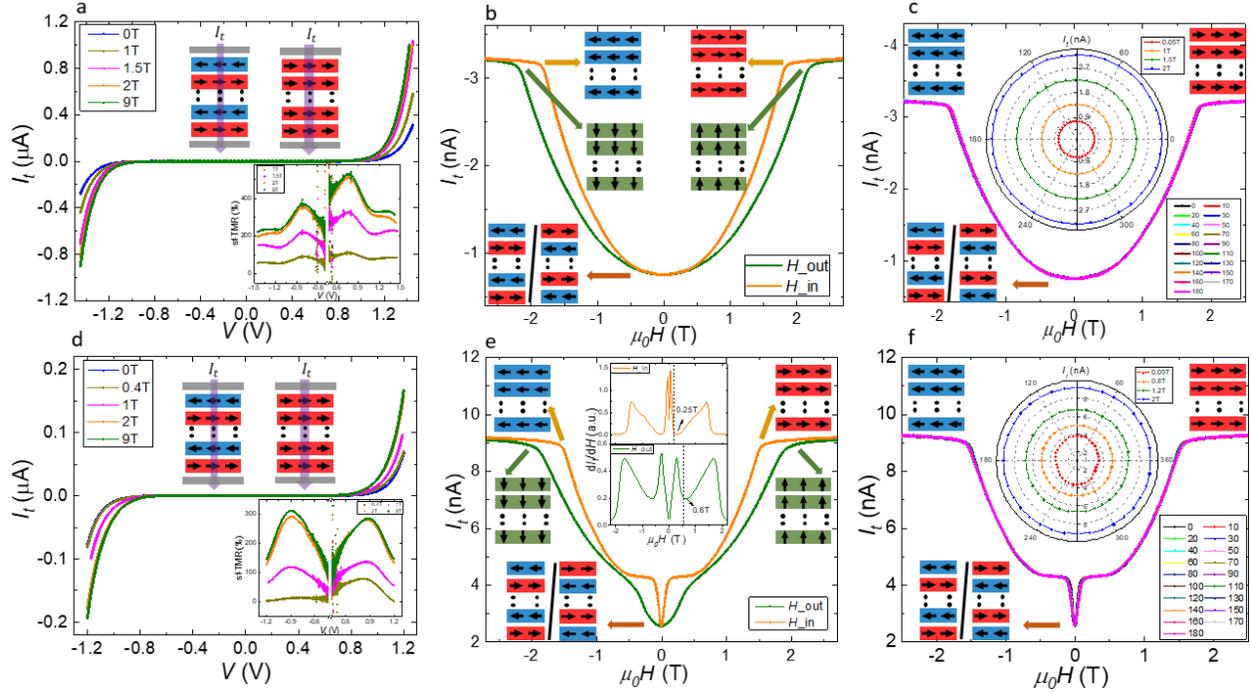

**Figure S5.** Electron tunneling through multilayer CrCl$_3$ sf-MTJs. Characterizations on two devices are shown in (a-c) and (d-f) respectively. (a, d) Tunneling current at selected in-plane magnetic fields. Top inset: Schematic of the magnetic configuration for $\mu_0 H_{in}$ = 0 T and 9 T. Bottom inset: Extracted sf-TMR ratio as a function of the bias voltage. (b, e) Tunneling current as a function of the in-plane (orange) and out-of-plane (green) magnetic field at selected DC bias voltage $V_{dc}$ = -900 mV (b) and $V_{dc}$ = 900 mV (e). Insets show the corresponding magnetic states. The top middle inset of (e) shows the derivative of the tunneling current to the in-plane (orange) and out-of-plane (green) magnetic field extracted from (e). (c, f) Tunneling current as a function of the in-plane magnetic field at selected angles $\theta$ between the field and the device. $V_{dc}$ = -900 mV (c) and $V_{dc}$ = 900 mV (f). Insets show the corresponding magnetic states. Top middle inset: Polar plot of $|I_t|$ as a functions of $\theta$ at selected magnetic fields. Data points extracted from (c, f).

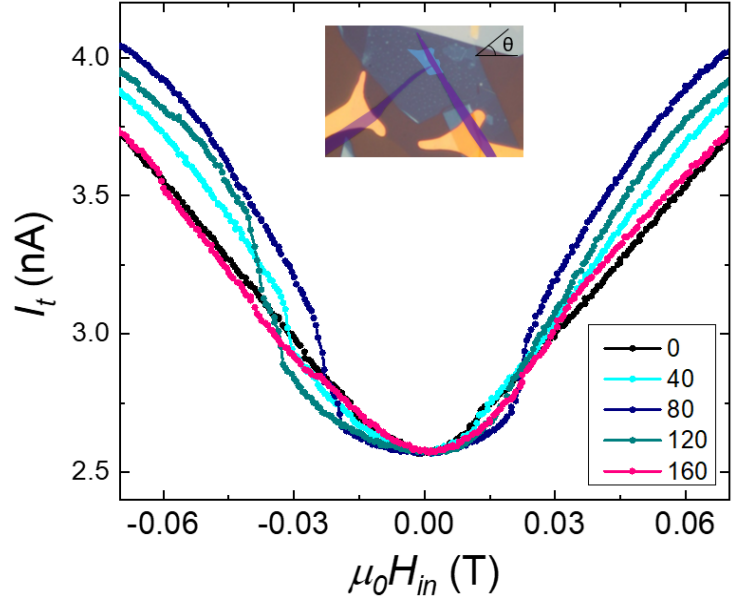

**Figure S6.** Spin-flop transition in multilayer $CrCl_3$. Tunneling current as a function of the in-plane magnetic field at selected angles $\theta$ between the field and a multilayer $CrCl_3$ sf-MTJ. $V_{dc}$ = 900 mV. Inset: false color optical microscope image of the device with the defined $\theta$.

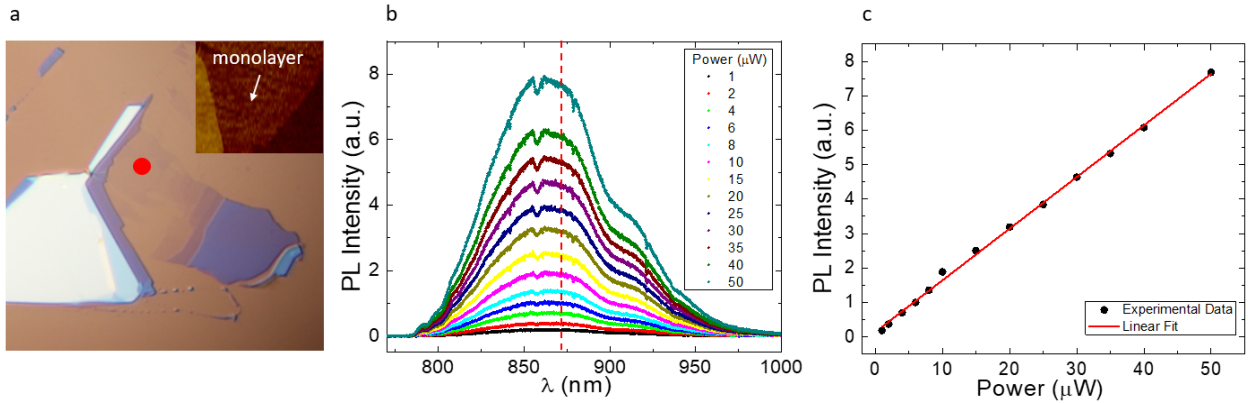

**Figure S7.** Power dependent PL of monolayer $CrCl_3$. (a) Optical microscope image of a $CrCl_3$ flake. The red spot represents the laser beam spot on the monolayer part. Inset: Atomic force micrograph of another monolayer $CrCl_3$ flake. (b) PL spectrum from the monolayer $CrCl_3$ sample shown in (a) with linearly polarized excitation at selected laser powers. The dip in the middle of spectra is due to bad pixels in the CCD. (c) PL intensity at 870 nm as a function of the excitation power. Data points extracted from (b). The red line shows a linear fit to the data.

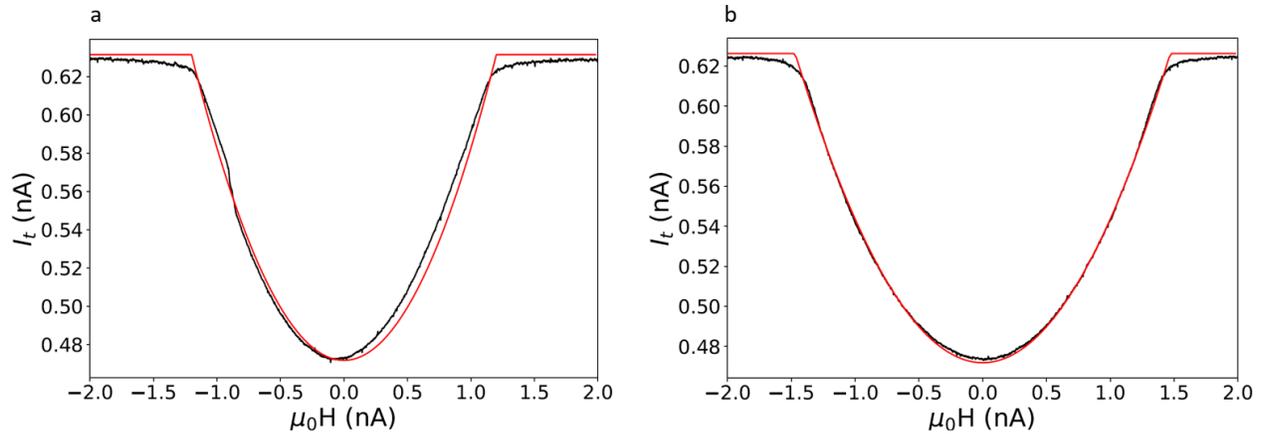

**Figure S8.** The numerical fitting for the tunneling current under the in-plane (a) and the out-of-plane (b) magnetic field. The red/black curve is the numerical/experimental result, respectively.